\documentclass[prl,twocolumn,showpacs]{revtex4-1}
\usepackage[utf8x]{inputenc}
\usepackage{amsbsy}
\usepackage{graphicx}
\usepackage{color}
\usepackage{dcolumn}
\usepackage{amsmath}
\usepackage{amssymb}
\usepackage{amsfonts}
\usepackage{subfigure}

\usepackage[colorlinks=true, citecolor=black, linkcolor=black, urlcolor=black]{hyperref}

\newcommand{\abinitio}{\emph{ab initio}}
\newcommand{\etal}{\emph{et al.}}

\newcommand{\bra}[1]{\langle #1|}
\newcommand{\ket}[1]{|#1 \rangle }
\newcommand{\beq}{\begin{equation}}
\newcommand{\eeq}{\end{equation}}
\newcommand{\beqn}{\begin{eqnarray}}
\newcommand{\eeqn}{\end{eqnarray}}

\begin{document}
\title{
First-Principles Theory of Anharmonicity and the
Inverse Isotope Effect in Superconducting Palladium-Hydride Compounds
}

\author{Ion Errea$^{1, 2}$}
\author{Matteo Calandra$^{1}$}
\author{Francesco Mauri$^{1}$}

\affiliation{$^1$Universit\'e Pierre et Marie Curie, CNRS, IMPMC, 4 Place Jussieu, 75252 Paris,  France}
\affiliation{$^2$IKERBASQUE, Basque Foundation for Science, 48011, Bilbao, Spain}

\begin{abstract}
Palladium hydrides display the largest isotope effect anomaly known
in literature. Replacement of hydrogen with the heavier
isotopes leads to higher superconducting temperatures, a
behavior inconsistent with harmonic theory.  
Solving the self-consistent harmonic
approximation by a stochastic approach, we obtain the anharmonic free
energy, the thermal expansion and the superconducting
properties fully \abinitio.
We find that the phonon spectra 
are strongly renormalized by anharmonicity far beyond the
perturbative regime. Superconductivity is phonon
mediated, but the harmonic approximation largely
overestimates the superconducting critical temperatures.
We explain the inverse isotope effect, obtaining a -0.38
value for the isotope coefficient in good agreement with experiments,
hydrogen anharmonicity being the main responsible for the isotope anomaly.

\end{abstract}


\maketitle

%
%

The explanation of the ion-mass isotope effect in phonon-mediated
superconductors  
is one of the greatest success of 
BCS theory \cite{PhysRev.108.1175}.
In a BCS superconductor composed of only one
type of ions of mass $M$,  the superconducting critical temperature
($T_c$) is expected to behave as $T_c\propto M^{-\alpha}$,
where $\alpha=0.5$ is the isotope coefficient. In conventional
superconductors with more atomic species, the 
total isotope coefficient should also be close to $0.5$.
However, in many superconductors like MgB$_2$ 
\cite{Hinks},  
fullerides 
\cite{PhysRevLett.83.404}
or high-$T_c$ cuprates 
\cite{hinks457,PhysRevLett.58.2333,liu64} the
isotope coefficient is substantially reduced and, 
in the most extreme case of palladium
hydrides (PH), it is even negative~\cite{stritzker257,PhysRevB.10.3818,
Schirber1984837}.

An isotope coefficient 
$\alpha=0.5$ relies on the following assumptions:
(i) the phonon frequencies are harmonic, consequently,
(ii) the electron-phonon interaction is mass independent, 
and (iii) the electron-electron interaction is not affected by the
isotope substitution.  Thus, a reduced isotope effect can either be
the fingerprint of a non-conventional mechanism
(e.g. spin-fluctuations or correlated  superconductivity)
or the breakdown of one of these assumptions
(e.g. anharmonicity). In both cases, the
superconducting state is considered anomalous and current 
state-of-the-art calculations do not quantitatively 
account for the behavior of
$T_c$ as a function of the isotope mass. This is due to
the difficulties of dealing either with non-conventional 
mechanisms or with anharmonicity.

Here we consider the most pathological case present in literature, 
the inverse isotope effect in PH. PdH has 
$T_c=8-9$K~\cite{stritzker257,PhysRevB.10.3818}.
Hydrogen substitution with the {\it heavier} deuterium leads
to a {\it higher} $T_c$, as $T_c$(PdD) 
$\approx 10-11$K~\cite{stritzker257,PhysRevB.10.3818}, leading to
$\alpha = -[\mathrm{ln} T_c(\mathrm{PdD}) - \mathrm{ln} T_c(\mathrm{PdH})]
/\mathrm{ln}2 \approx -0.3$. Remarkably, PdT has a
higher $T_c$ than PdD, but there is no experimental value at full
stoichiometry~\cite{Schirber1984837}.
A considerable theoretical and experimental effort 
~\cite{karakozov681,Kolesnikov1991257,
Sherman1977353,PhysRevB.58.2591,
Chowdhury1973229,
PhysRevLett.33.1297,PhysRevLett.57.2955,
PhysRevLett.34.144,PhysRevLett.39.574,
PhysRevB.44.10377,PhysRevB.45.12405,PhysRevB.12.117,PhysRevB.29.4140,
karakozov681}
has been devoted to
explain this phenomenon over the last decades.
Karakozov \etal{}~\cite{karakozov681}, in a pioneering work,
studied anharmoniciy in PH in the framework of perturbation theory
to the bare harmonic phonon frequency and concluded that the negative
isotope effect could be due to anharmonicity, as other
authors suggested later~\cite{PhysRevB.45.12405}. 
Other explanations based on electronic properties~\cite{PhysRevLett.34.144},
zero-point motion~\cite{PhysRevB.29.4140} and volume effects~\cite{PhysRevB.12.117}
have been invoked as well.
Inelastic neutron scattering
experiments~\cite{PhysRevLett.33.1297} show strongly temperature
dependent phonon lifetimes, a clear fingerprint of 
anharmonicity. Furthermore, \abinitio{} total energy 
calculations~\cite{PhysRevB.44.10377,PhysRevB.45.12405}
suggest that the potential felt by the hydrogen atoms is very anharmonic.
Nevertheless, no state-of-the-art calculations of the
electron-phonon interaction and anharmonicity are present
so that the interplay of these effects is still unclear.

In this work we study the occurrence of phonon-mediated
superconductivity in PH and show that the inverse isotope 
effect is quantitatively explained by the inclusion of anharmonicity. 
Anharmonicity is so large that perturbative approaches 
\cite{karakozov681,PhysRevB.82.104504,PhysRevB.87.214303, Lazzeri} are not feasible in PH.
To solve this issue, we implement
the self-consistent harmonic approximation (SCHA)
\cite{hooton422} within a first-principles approach.
Differently from other methods developed to deal with anharmonic
effects  \cite{PhysRevLett.100.095901,PhysRevB.84.180301,Needs}, our method allows to
access directly the anharmonic free energy of the system,
with full inclusion of the anharmonic potential terms, 
and is variational
in the free energy with respect to a trial harmonic potential.
Moreover, compared to other 
implementations of the SCHA \cite{PhysRevLett.106.165501,
errea:112604}, we
replace the cumbersome calculation of anharmonic coefficients 
by the evaluation of atomic forces
on supercells with suitably chosen stochastic ionic configurations.  

%
%

The ionic Hamiltonian is ${H} = {T} + {V}$, where 
${T}$ and ${V}$ are the kinetic and potential energy
operators. In the adiabatic approximation
the potential is defined by the Born-Oppenheimer (BO) energy surface.
Then, the free energy of the ionic system can be calculated as
$F_H = - \frac{1}{\beta} \ln Z_H$, where the partition function
is $Z_H = \mathrm{tr} [e^{-\beta {H}}]$ and 
$\beta= 1 / ( k_B T ) $.
A quantum variational principle in the free energy can 
be established for an arbitrary
trial Hamiltonian ${\mathcal{H}}={T} +{\mathcal{V}}$ as 
\cite{hooton422}
\beq
F_H \le \mathcal{F}_H[\mathcal{H}] = F_{\mathcal{H}}
         + \int \mathrm{d}\mathbf{R}[V(\mathbf{R}) - \mathcal{V}(\mathbf{R})] 
        \rho_{\mathcal{H}}(\mathbf{R})
,
\label{gibbs-bogoliubov}
\eeq
where 
$\rho_{\mathcal{H}}(\mathbf{R}) = 
\bra{\mathbf{R}} e^{-\beta{\mathcal{H}}}
\ket{\mathbf{R}} / Z_{\mathcal{H}}$ is the probability to find a system described by
${\mathcal{H}}$ in a general $\mathbf{R}$
ionic configuration. The equality holds for ${\mathcal{H}}={H}$.
The SCHA takes a harmonic ${\mathcal{V}}$ and
minimizes $\mathcal{F}_H[\mathcal{H}]$ with respect to it~\cite{hooton422}.
One advantage of taking a harmonic potential is that
$F_{\mathcal{H}}$ and $\rho_{\mathcal{H}}(\mathbf{R})$ can be expressed in a closed form 
in terms of the phonon frequencies and polarizations~\cite{doi:10.1021/jp044251w}.
In particular, $\rho_{\mathcal{H}}(\mathbf{R})$ is Gaussian and is given as
\beq
\rho_{\mathcal{H}}(\mathbf{R}) = A_{\mathcal{H}} 
     \exp \left[ - \sum\limits_{st \alpha \beta \mu} \frac{ \sqrt{M_sM_t} }{ 
     2 a_{\mu \mathcal{H}}^2}
 \epsilon_{\mu \mathcal{H}}^{s\alpha}
                \epsilon_{\mu \mathcal{H}}^{t\beta} u^{s\alpha} u^{t\beta} 
    \right],
\label{probability}
\eeq
where $A_{\mathcal{H}}$ is the normalization constant,
$s$ and $t$ are atom indexes, $\alpha$ and $\beta$ Cartesian indexes, 
$\mu$ is a mode index, 
$M$ denotes the mass of an atom, 
$\mathbf{u} = \mathbf{R} - \mathbf{R}_{\mathrm{eq}}$ is the
displacement from equilibrium, 
$a^2_{\mu \mathcal{H}}=\hbar \coth (\beta \hbar \omega_{\mu \mathcal{H}} / 2) 
/ ( 2 \omega_{\mu \mathcal{H}} ) $,
and $\{\omega_{\mu \mathcal{H}}\}$ and 
$\{ \epsilon_{\mu \mathcal{H}}^{s\alpha} \}$ 
represent the phonon frequencies and polarizations defined by
${\mathcal{H}}$.

As long as the equilibrium positions are fixed
by symmetry, as in PH, minimizing $\mathcal{F}_H[\mathcal{H}]$ 
with respect to ${\mathcal{V}}$ is equivalent
to performing the minimization with respect to the force constant matrix
$C$, which defines the trial harmonic potential as 
$\mathcal{V}=\frac{1}{2}\sum_{st \alpha \beta}  u^{s\alpha}
C_{st}^{\alpha \beta} u^{t\beta}$. The minimization 
is carried through the components
of $C$ in the basis of $N \times N$ hermitian matrices 
preserving crystal symmetries, where $N$ is the number of modes.
We minimize $\mathcal{F}_H[\mathcal{H}]$ by a conjugate-gradient 
(CG) algorithm, which requires the knowledge of the gradient 
$\boldsymbol{\nabla} \mathcal{F}_H[\mathcal{H}]$ with respect
to $C$:
\beq
\boldsymbol{\nabla} \mathcal{F}_H[\mathcal{H}]  =  
  - \sum_{st\alpha \beta} 
   E^{st \alpha \beta}_{\mathcal{H}}  
   \int \mathrm{d}\mathbf{R}
        \tilde{f}_{\mathcal{H}}^{s\alpha}(\mathbf{R}) u^{t\beta}
        \rho_{\mathcal{H}}(\mathbf{R}),
\label{gradient-no-stat}
\eeq
where $E^{st \alpha \beta}_{\mathcal{H}} = \sum_{\mu}
\sqrt{M_t/M_s} ( \epsilon_{\mu \mathcal{H}}^{s \alpha} 
    \boldsymbol{\nabla} \ln a_{\mu \mathcal{H}}
   + \boldsymbol{\nabla} \epsilon_{\mu \mathcal{H}}^{s \alpha} 
     ) \epsilon_{\mu \mathcal{H}}^{t \beta}$, and
$\tilde{f}_{\mathcal{H}}^{s\alpha}(\mathbf{R}) = 
f^{s \alpha}(\mathbf{R}) - f_{\mathcal{H}}^{s \alpha}(\mathbf{R})$ 
is the difference between the force
on the $s$-th atom along the direction $\alpha$,
$f^{s \alpha}(\mathbf{R})$,  
and  the harmonic force derived from ${\mathcal{V}}$, 
$f_{\mathcal{H}}^{s \alpha}(\mathbf{R})$.
At each step $j$ of the CG
minimization, the trial harmonic Hamiltonian
is updated to ${\mathcal{H}}_j$, till the minimum is found. At the minimum,
the $\{\omega_{\mu \mathcal{H}}\}$ frequencies 
form the SCHA phonon spectrum renormalized by anharmonicity

In our stochastic SCHA (SSCHA) approach, both $\mathcal{F}_H[\mathcal{H}]$ 
and $\boldsymbol{\nabla} \mathcal{F}_H[\mathcal{H}]$ 
are calculated making use of importance sampling and
reweighting techniques. 
We start defining an initial trial ${\mathcal{H}}_0$ 
harmonic Hamiltonian 
and creating a set of $\{\mathbf{R}_I\}_{I=1,\dots,N_c}$ ionic configurations 
in a supercell according to the $\rho_{\mathcal{H}_0}(\mathbf{R})$ 
distribution given in Eq. \eqref{probability}. 
These configurations are trivially created making use of 
random numbers generated with a Gaussian distribution.
Secondly, we calculate the BO energy and the atomic
forces for each random configuration $\mathbf{R}_I$,
$V(\mathbf{R}_I)$ and $f^{s\alpha}(\mathbf{R}_I)$, respectively.
This allows us to evaluate the integrals 
in Eqs. \eqref{gibbs-bogoliubov}
and \eqref{gradient-no-stat} as an average of the integrands 
over the $N_c$ configurations 
(importance sampling). Thus, we can compute the 
free energy and its gradient, and perform the first
CG step to obtain ${\mathcal{H}}_1$.
In principle we should reevaluate BO energies and forces for
the supercell at each CG step $j$, a very time-demanding 
task as generally hundreds of steps are needed to converge.
This can be avoided with a reweighting procedure. We introduce 
the reweighting $\rho_{\mathcal{H}_j}(\mathbf{R}_I)/
\rho_{\mathcal{H}_0}(\mathbf{R}_I)$ factor (equal to one
in the first $j=0$ step) in the importance sampling evaluation of the integrals.
Namely, $\mathcal{F}_H[\mathcal{H}_j]$ and 
$\boldsymbol{\nabla} \mathcal{F}_H[\mathcal{H}_j]$ 
are obtained as:
\beq
\mathcal{F}_H[\mathcal{H}_j] \simeq F_{\mathcal{H}_j} +
        \frac{1}{N_c} \sum_{I=1}^{N_c} [V(\mathbf{R}_I) - \mathcal{V}_j(\mathbf{R}_I)] 
         \frac{\rho_{\mathcal{H}_j}(\mathbf{R}_I)}{
               \rho_{\mathcal{H}_0}(\mathbf{R}_I)},
\label{potentialintegral}
\eeq
\beq
\boldsymbol{\nabla} \mathcal{F}_H[\mathcal{H}_j]  \simeq  
  - \sum_{st\alpha \beta} E^{st \alpha \beta}_{\mathcal{H}_j}  
    \frac{1}{N_c} \sum_{I=1}^{N_c}  
        \tilde{f}_{\mathcal{H}_j}^{s\alpha}(\mathbf{R}_I) u_{I}^{t\beta}
               \frac{\rho_{\mathcal{H}_j}(\mathbf{R}_I)}{
               \rho_{\mathcal{H}_0}(\mathbf{R}_I)},
\label{gradient}
\eeq 
where the equality holds for  $N_c \to \infty$.
Including the reweighting factor, we can use the BO energies and forces 
of the configurations created with the initial $\rho_{\mathcal{H}_0}(\mathbf{R})$
distribution also for the following $j$ CG iterations.
However, if $\frac{1}{N_c}\sum_{i=1}^{N_c} 
\rho_{\mathcal{H}_j}(\mathbf{R}_i) /
\rho_{\mathcal{H}_0}(\mathbf{R}_i)$ deviates 
substantially from one, 
$\mathcal{F}_H[\mathcal{H}_j]$ and 
$\boldsymbol{\nabla} \mathcal{F}_H[\mathcal{H}_j]$
cannot be accurately evaluated
anymore as the initial set of configurations
does not represent closely 
$\rho_{\mathcal{H}_j}(\mathbf{R})$. When this occurs,
we use the probability distribution of the
current step, $\rho_{\mathcal{H}_j}(\mathbf{R})$,
to create a new set of configurations for which we recompute
atomic forces and BO energies to be used
in the present and subsequent CG iterations.
The process continues till the gradient vanishes.

\begin{figure}[t]
\includegraphics[width=0.46\textwidth]{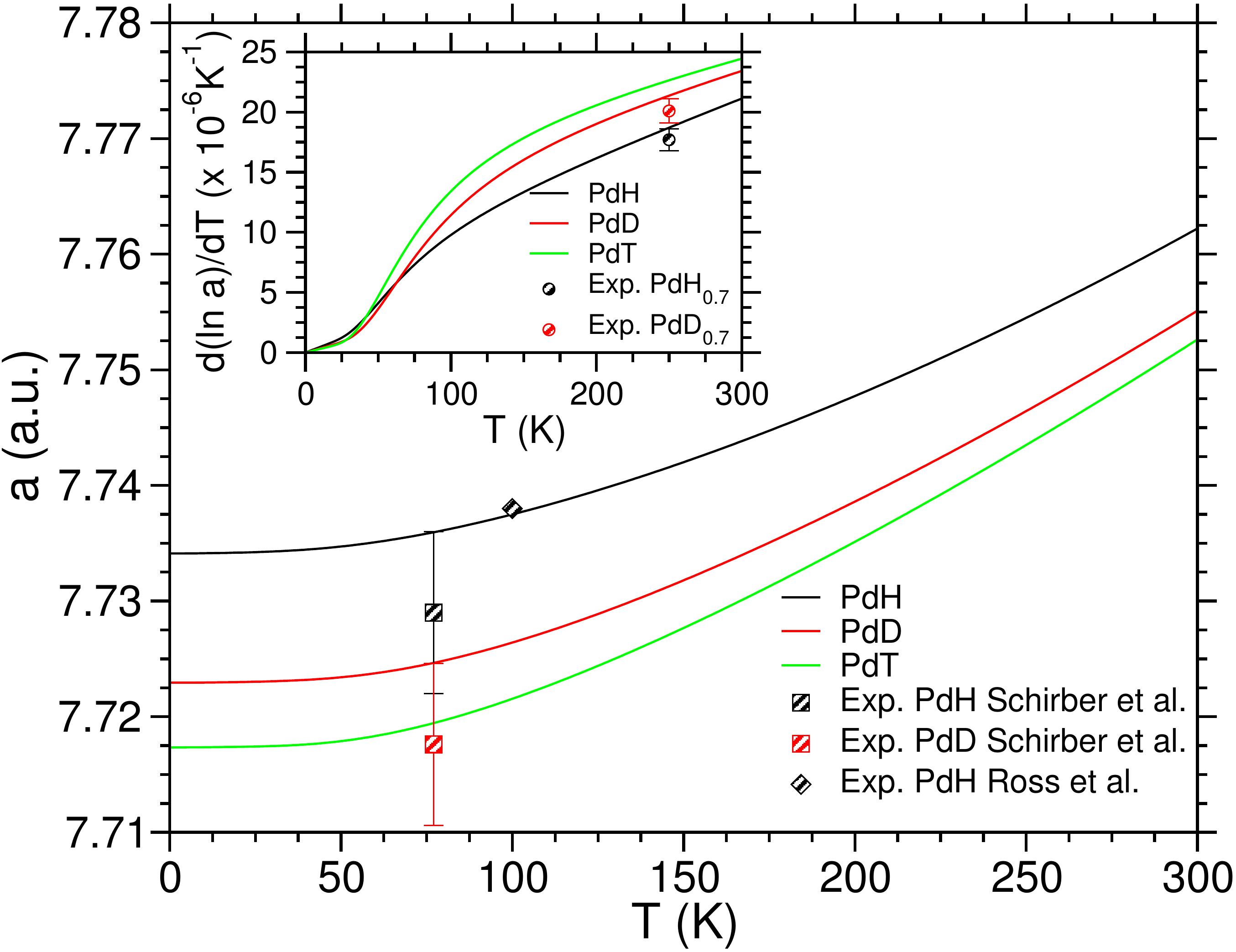}
\caption{(Color online) 
Equilibrium lattice 
parameters as a function of temperature for 
PdH, PdD and PdT compared to experimental 
results~\cite{PhysRevB.12.117,PhysRevB.58.2591}. In the inset 
the calculated thermal expansion coefficients are shown together
with the measured values in Ref.~\cite{0305-4608-10-3-006}.
} 
\label{fig1}
\end{figure}

\begin{figure}[t]
\includegraphics[width=0.41\textwidth]{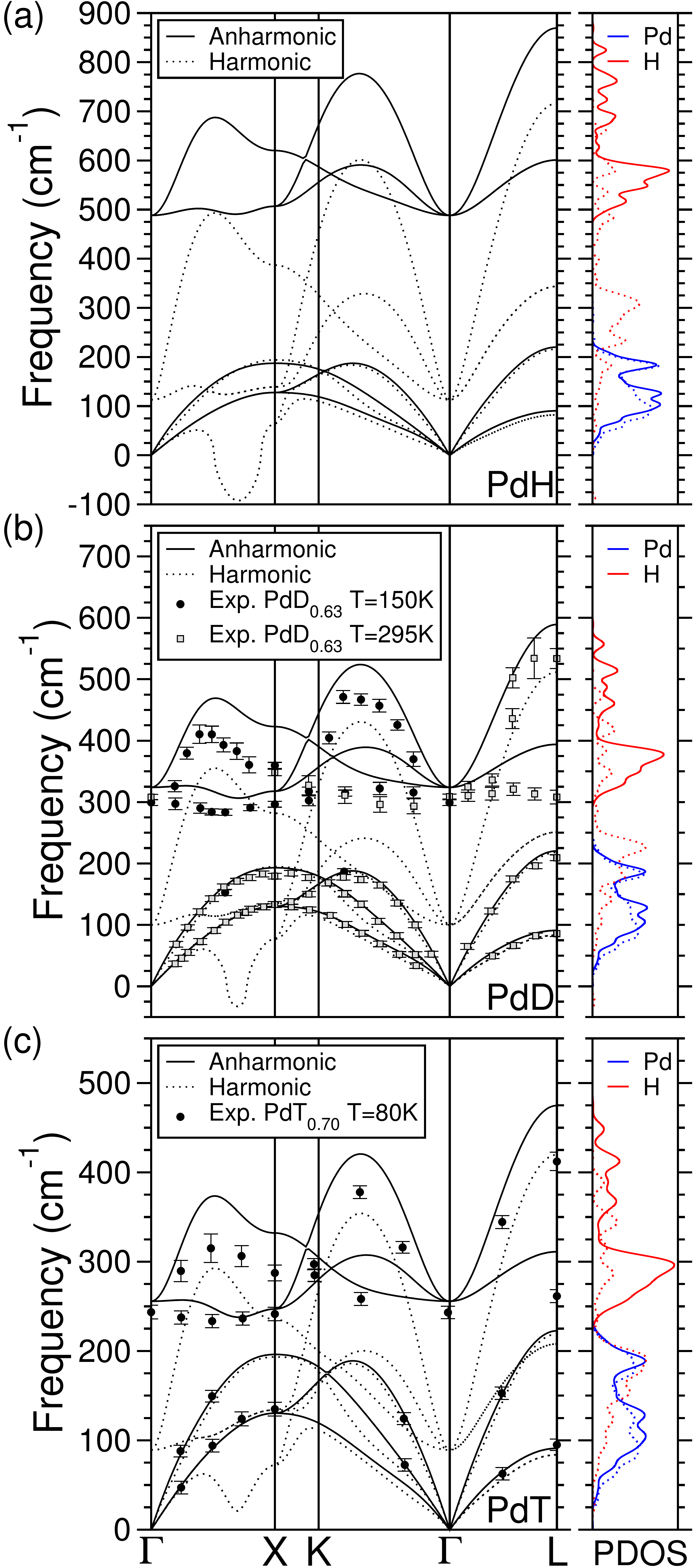}
\caption{(Color online) Harmonic and SSCHA phonon spectra calculated for PdH (a),
PdD (b) and PdT (c) at the equilibrium volume at 0 K.
The experimental results obtained for 
non-stoichiometric PdD$_{0.63}$~\cite{PhysRevLett.33.1297} and
PdT$_{0.70}$~\cite{PhysRevLett.57.2955} are shown.
In the right panels the PDOS 
projected onto Pd and H atoms is
plotted within the harmonic approximation (dotted lines) and
the SSCHA (solid lines).}
\label{fig2}
\end{figure}

We apply this method to stoichiometric PdH, PdD and PdT.
Total energies, atomic forces,
harmonic phonons and deformation potentials needed for the 
Eliashberg functions are computed
with density-functional theory (DFT) and linear 
response~\cite{0953-8984-21-39-395502,RevModPhys.73.515}. We use
the Perdew-Zunger local-density approximation~\cite{PhysRevB.23.5048}
and ultrasoft pseudopotentials~\cite{Note1}.

%
%

We first demonstrate the capability of our developed method to obtain
the free energy as a function of temperature. We calculate 
$\mathcal{F}_H[\mathcal{H}]$ and 
$\boldsymbol{\nabla} \mathcal{F}_H[\mathcal{H}]$
from 
a model potential combining  the \abinitio{} harmonic potential with a 
fourth-order on-site anharmonic potential
fitted to the DFT total energies~\cite{footnote}. 
The potential satisfies the 
symmetries of the rock-salt structure~\cite{PhysRevB.44.10377}
and has also been used to study cubic ferroelectrics~\cite{PhysRevB.52.6301}. 
In this system anharmonic forces are short-range 
and, consequently, phonon dispersions calculated with the model potential
and with \abinitio{} forces are in close agreement (see Supplemental Material). 
The use of the model allows us to estimate the free energy at several 
volumes and temperatures very efficiently. 
Minimizing the free energy we obtain the lattice parameter of
each isotope as a function of temperature.
The absolute values of the 
lattice parameters and the thermal expansion coefficients, Fig. \ref{fig1}, are in good
agreement with measurements~\cite{PhysRevB.12.117,PhysRevB.58.2591,
0305-4608-10-3-006}.

Once the equilibrium volumes are determined,
we obtain the SSCHA phonon dispersions computing the
forces on the $\{\mathbf{R}_I\}_{I=1,\dots,N_c}$
configurations completely from first principles~\cite{footnote}, overcoming the approximation of a model potential.
%
The results are shown in Fig. \ref{fig2} and compared with available
experimental data on deuterium and tritium deficient samples.
The breakdown of the harmonic approximation is evident in all systems, 
particularly in PdH and PdD displaying imaginary phonon frequencies.
The anharmonic correction given by the SSCHA is larger
than the harmonic phonon frequency itself, invalidating any possible
perturbative approach. Interestingly, both the low-energy acoustic and high-energy
optical modes are affected by anharmonicity even if the largest
correction involves the H-character optical modes.
At zone center the PdH optical
modes are degenerate at 488 cm$^{-1}$, in good agreement   
with 
inelastic neutron and Raman experiments 
~\cite{Sherman1977353,
Kolesnikov1991257,Chowdhury1973229,PhysRevB.58.2591}
(around 450-472 cm$^{-1}$ at different temperatures
and H concentrations).
Since the H atom is smaller than the octahedral void
in the Pd fcc lattice, 
hydrogen vibrations are characterized by 
very anharmonic rattling modes.
This is confirmed by the weak mass dependence
of the root-mean square displacement of 
hydrogen (0.55 a.u. in PdH, 0.48 a.u. in PdD and
0.44 a.u. in PdT at 0 K), which does not scale according to the
harmonic $M^{-0.5}$ relation.  
  
\begin{figure}[t]
\includegraphics[width=0.46\textwidth]{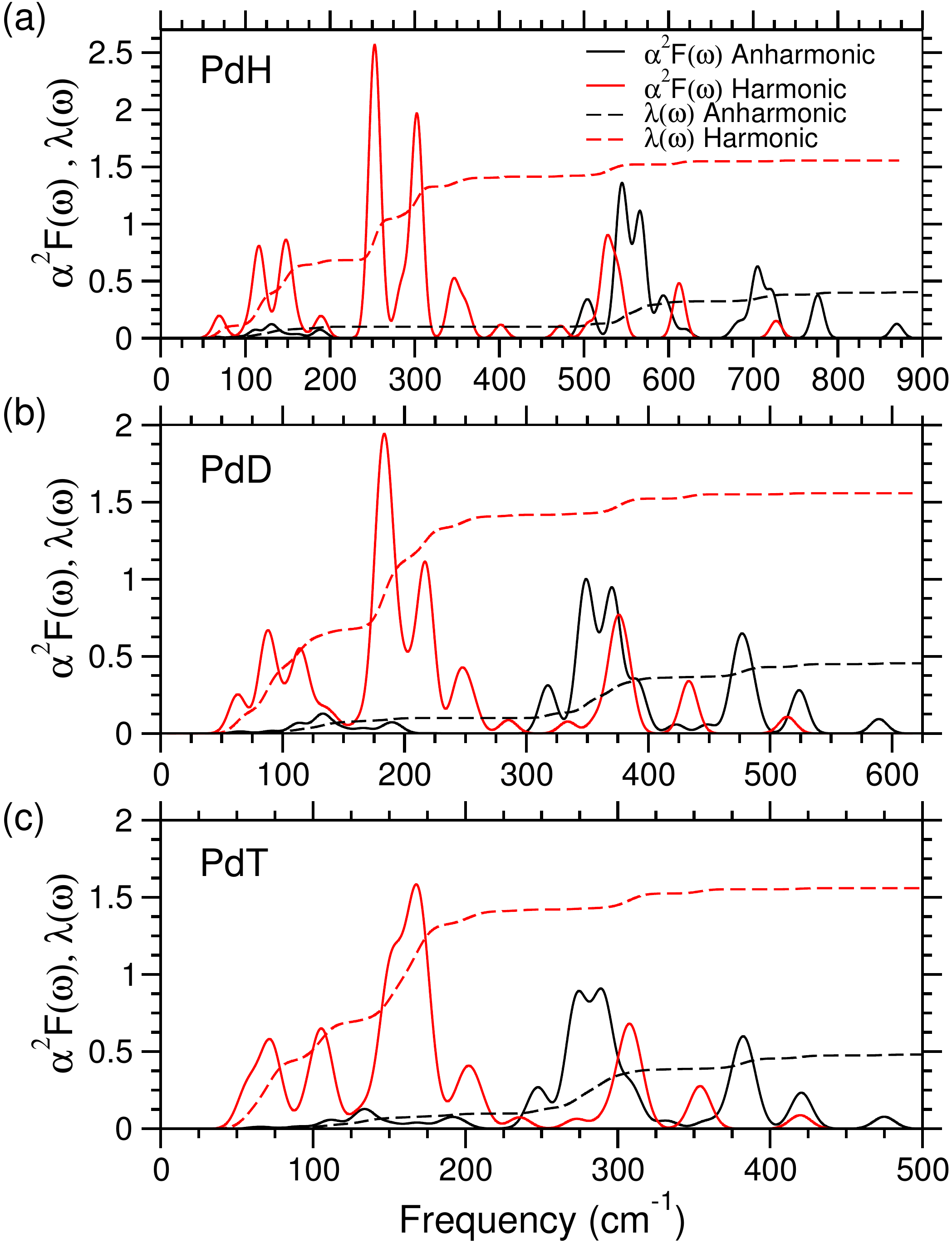}
\caption{ (Color online) $\alpha^2F(\omega)$ and 
$\lambda(\omega) = 2 \int_0^{\omega} \mathrm{d}\omega' \alpha^2F(\omega')/\omega' $
for (a) PdH, (b) PdD and (c) PdT within the harmonic approximation 
and the SSCHA.} 
\label{fig3}
\end{figure}

From the calculated phonon spectra we obtain the Eliashberg function
as
\beqn
&& \alpha^{2}F(\omega) = \frac{1}{N(0) N_k N_q} \sum_{{\bf k}{\bf q}nm} 
  \sum_{st \alpha \beta \mu} 
  \frac{\epsilon_{\mu}^{s \alpha}(\mathbf{q}) 
           \epsilon_{\mu}^{t \beta *}(\mathbf{q}) }{
        2 \omega_{\mu} (\mathbf{q}) \sqrt{M_sM_t}} \nonumber \\
&& \ \times    
   d^{s\alpha}_{{\bf k}n,{\bf k}+{\bf q}m}
   d^{t\beta *}_{{\bf k}n,{\bf k}+{\bf q}m}
   \delta(\epsilon_{{\bf k}n}) 
   \delta(\epsilon_{{\bf k+q}m})
   \delta(\omega -  \omega_{\mu} (\mathbf{q})),
\label{eliashberg}
\eeqn
where $d^{s\alpha}_{{\bf k}n,{\bf k}+{\bf q}m} = \bra{{\bf k}n}
\delta V / \delta u^{s\alpha}(\mathbf{q}) \ket{{\bf k}+{\bf q}m}$
is the deformation potential, $\ket{{\bf k}n}$ is a  
Kohn-Sham state with energy $\epsilon_{{\bf k}n}$ 
measured from the Fermi  level ($\epsilon_F$), 
$N_k$ and $N_q$ are the number of electron
and phonon momentum points used for the Brillouin-zone (BZ) sampling, 
and $N(0)$ is the density of states per spin at $\epsilon_F$. 
We compute $\alpha^{2}F(\omega)$
in the
harmonic or anharmonic case by using the harmonic or SSCHA
phonon frequencies and polarizations  
in Eq. \eqref{eliashberg}. The electron-phonon coupling
constant $\lambda$, as well as the logarithmic frequency average
$\omega_{\mathrm{log}}$, 
are obtained as
$\lambda = 2 \int_0^{\infty} \mathrm{d}\omega \alpha^2F(\omega)/\omega$
and
$\omega_{\mathrm{log}} = \exp( \frac{2}{\lambda} 
\int_0^{\infty} \mathrm{d}\omega \alpha^2F(\omega)
\ln \omega /\omega)$
~\cite{PhysRevB.12.905}. 
We estimate $T_c$ from
the solution of the single-band Migdal-Eliashberg 
equations, using $\mu^*=0.085$ as 
calculated in Ref.~\cite{PhysRevLett.39.574}.
In the harmonic approximation the  
equilibrium volume of PdT is used for all isotopes as, in this 
case, there are no imaginary phonons. 
$\alpha^{2}F(\omega)$  functions are shown in
Fig. \ref{fig3} and the results for $T_c$ are
presented in Table \ref{table1}.

\begin{table}[b]
\caption{Calculated $\lambda$, $\omega_{\mathrm{log}}$ and $T_c$ values for PH.
$\omega_{\mathrm{log}}$ values are given in 
cm$^{-1}$ and $T_c$'s in K. Experimental values for 
$T_c$~\cite{stritzker257,PhysRevB.10.3818} are presented as well. 
The value of $\alpha$ between isotopes PdA and PdB is
calculated as $\alpha_{\mathrm{PdA(B)}} = 
-(\mathrm{ln} T_c(\mathrm{PdB}) - \mathrm{ln} T_c(\mathrm{PdA}))
/(\mathrm{ln}M_B - \mathrm{ln}M_A)$.
}
\begin{center}
\begin{tabular*}{0.43\textwidth}{c | c c c | c c c | c c } 
\hline
\hline
& \multicolumn{3}{c | }{Harmonic} & \multicolumn{3}{c | }{SSCHA} & \multicolumn{2}{c }{Exp.}\\
& $\lambda$ & $\omega_{\mathrm{log}}$ & $T_c$ & $\lambda$ & $\omega_{\mathrm{log}}$ & $T_c$ 
& $T_c$~\cite{stritzker257} & $T_c$~\cite{PhysRevB.10.3818} \\   
\hline
PdH    & 1.55 & 205 & 47  & 0.40 & 405 & 5.0 & 9  & 8  \\ 
PdD    & 1.55 & 150 & 34  & 0.46 & 304 & 6.5 & 11 & 10 \\
PdT    & 1.55 & 125 & 30  & 0.48 & 257 & 6.9 \\ 
\hline
$\alpha_{\mathrm{PdH(D)}}$ &  &  & 0.47 & & & -0.38 & -0.29 & -0.32 \\ 
$\alpha_{\mathrm{PdH(T)}}$ &  &  & 0.41 & & & -0.29 &       &       \\ 
\hline
\hline
\end{tabular*}
\end{center}
\label{table1}
\end{table}

In the harmonic approximation $\lambda$ is independent of the mass and the 
isotope coefficient is determined by $\omega_{\mathrm{log}}$. Thus, the 
heavier the isotope the lower $T_c$ and $\alpha$
is close to 0.5. The harmonic approximation strongly 
overestimates $T_c$, predicting very high values for all compounds. 
This overestimation is due to the very soft H-character
vibrations that contribute as $\lambda \propto 
\omega^{-2}$
to the electron-phonon coupling,
leading to extraordinary values for $\lambda$.
This is evident from the comparison between the 
phonon density of states (PDOS), Fig. \ref{fig2},
and $\alpha^{2}F(\omega)$, Fig. \ref{fig3}.  

In the SSCHA, 
both $\lambda$ and, consequently, $T_c$ are substantially reduced
due to the enhancement of the frequencies induced by anharmonicity. 
The hardening is more important the lighter the isotope
as lighter atoms feel the potential
farther away from equilibrium because of the fluctuations 
of the zero-point motion.
This makes $\lambda$ mass-dependent and
larger the heavier the isotope. Remarkably, the differences in $\lambda$ explain the
inverse isotope effect in $T_c$ and the value we obtain for the isotope coefficient
$\alpha$ is in good agreement with experiments
(see Table \ref{table1}). The obtained $T_c$'s are close to 
experimental results even if no anharmonic corrections were
incorporated into the deformation potential.
As noted by $\alpha^2F(\omega)$ and the integrated electron-phonon coupling 
$\lambda(\omega)$ in Fig. \ref{fig3}, 
H-character optical modes 
have the largest contribution to $\lambda$, between 75\% and 79\% of the
total depending on the isotope. The contribution of the 
low-energy Pd-character acoustic modes is similar for the
three hydrides. Thus, we conclude that
superconductivity and the inverse isotope effect 
in PH is driven by hydrogen anharmonicity. 

%
%

In summary, we present a stochastic implementation of the
SCHA that allows us to treat anharmonic effects in the 
non-perturbative regime. The method gives access directly to
the free energy of the system and is variational in the 
free energy with respect to a trial harmonic Hamiltonian. 
The method is applied to PH
calculating the free energy, the thermal
expansion, the anharmonic phonon spectra and the superconducting
properties. 
We demonstrate that superconductivity in PH is phonon-mediated
and the anomalous inverse isotope effect is due to the large
anharmonicity of hydrogen vibrations,
which is impossible to treat within perturbation theory. 
Our findings open 
new perspectives in the interpretation of reduced isotope effects
in superconductors.
Moreover, we demonstrate that anharmonicity induces 
a huge suppression of $T_c$, almost a factor of 10 in 
PdH. This poses the question of whether harmonic $T_c$
calculations in high-pressure metallic 
hydrides~\cite{Kim16022010} are
overestimated~\cite{PhysRevB.82.104504}, in particular, in the very similar 
PtH~\cite{PhysRevLett.107.117002}.
More generally, our methodological developments
will allow to investigate strongly anharmonic systems in the
non-perturbative regime, ranging from ferroelectrics~\cite{PhysRevB.52.6301}, 
charge-density wave systems~\cite{gruner}, and many more.

%
%

We are grateful to A. Bergara 
for fruitful discussions. IE acknowledges financial support
from the Department of Education, Language policy and Culture of the Basque
Government (Grant BFI-2011-65). Computer facilities 
were provided by CINES, CCRT and IDRIS.


\begin{thebibliography}{43}%
\makeatletter
\providecommand \@ifxundefined [1]{%
 \@ifx{#1\undefined}
}%
\providecommand \@ifnum [1]{%
 \ifnum #1\expandafter \@firstoftwo
 \else \expandafter \@secondoftwo
 \fi
}%
\providecommand \@ifx [1]{%
 \ifx #1\expandafter \@firstoftwo
 \else \expandafter \@secondoftwo
 \fi
}%
\providecommand \natexlab [1]{#1}%
\providecommand \enquote  [1]{``#1''}%
\providecommand \bibnamefont  [1]{#1}%
\providecommand \bibfnamefont [1]{#1}%
\providecommand \citenamefont [1]{#1}%
\providecommand \href@noop [0]{\@secondoftwo}%
\providecommand \href [0]{\begingroup \@sanitize@url \@href}%
\providecommand \@href[1]{\@@startlink{#1}\@@href}%
\providecommand \@@href[1]{\endgroup#1\@@endlink}%
\providecommand \@sanitize@url [0]{\catcode `\\12\catcode `\$12\catcode
  `\&12\catcode `\#12\catcode `\^12\catcode `\_12\catcode `\%12\relax}%
\providecommand \@@startlink[1]{}%
\providecommand \@@endlink[0]{}%
\providecommand \url  [0]{\begingroup\@sanitize@url \@url }%
\providecommand \@url [1]{\endgroup\@href {#1}{\urlprefix }}%
\providecommand \urlprefix  [0]{URL }%
\providecommand \Eprint [0]{\href }%
\providecommand \doibase [0]{http://dx.doi.org/}%
\providecommand \selectlanguage [0]{\@gobble}%
\providecommand \bibinfo  [0]{\@secondoftwo}%
\providecommand \bibfield  [0]{\@secondoftwo}%
\providecommand \translation [1]{[#1]}%
\providecommand \BibitemOpen [0]{}%
\providecommand \bibitemStop [0]{}%
\providecommand \bibitemNoStop [0]{.\EOS\space}%
\providecommand \EOS [0]{\spacefactor3000\relax}%
\providecommand \BibitemShut  [1]{\csname bibitem#1\endcsname}%
\let\auto@bib@innerbib\@empty
\bibitem [{\citenamefont {Bardeen}\ \emph {et~al.}(1957)\citenamefont
  {Bardeen}, \citenamefont {Cooper},\ and\ \citenamefont
  {Schrieffer}}]{PhysRev.108.1175}%
  \BibitemOpen
  \bibfield  {author} {\bibinfo {author} {\bibfnamefont {J.}~\bibnamefont
  {Bardeen}}, \bibinfo {author} {\bibfnamefont {L.~N.}\ \bibnamefont {Cooper}},
  \ and\ \bibinfo {author} {\bibfnamefont {J.~R.}\ \bibnamefont {Schrieffer}},\
  }\href {\doibase 10.1103/PhysRev.108.1175} {\bibfield  {journal} {\bibinfo
  {journal} {Phys. Rev.}\ }\textbf {\bibinfo {volume} {108}},\ \bibinfo {pages}
  {1175} (\bibinfo {year} {1957})}\BibitemShut {NoStop}%
\bibitem [{\citenamefont {Hinks}\ \emph
  {et~al.}(2001{\natexlab{a}})\citenamefont {Hinks}, \citenamefont {H},\ and\
  \citenamefont {Jorgensen}}]{Hinks}%
  \BibitemOpen
  \bibfield  {author} {\bibinfo {author} {\bibfnamefont {D.~G.}\ \bibnamefont
  {Hinks}}, \bibinfo {author} {\bibfnamefont {C.}~\bibnamefont {H}}, \ and\
  \bibinfo {author} {\bibfnamefont {J.~D.}\ \bibnamefont {Jorgensen}},\
  }\href@noop {} {\bibfield  {journal} {\bibinfo  {journal} {Nature}\ }\textbf
  {\bibinfo {volume} {411}},\ \bibinfo {pages} {457} (\bibinfo {year}
  {2001}{\natexlab{a}})}\BibitemShut {NoStop}%
\bibitem [{\citenamefont {Fuhrer}\ \emph {et~al.}(1999)\citenamefont {Fuhrer},
  \citenamefont {Cherrey}, \citenamefont {Zettl}, \citenamefont {Cohen},\ and\
  \citenamefont {Crespi}}]{PhysRevLett.83.404}%
  \BibitemOpen
  \bibfield  {author} {\bibinfo {author} {\bibfnamefont {M.~S.}\ \bibnamefont
  {Fuhrer}}, \bibinfo {author} {\bibfnamefont {K.}~\bibnamefont {Cherrey}},
  \bibinfo {author} {\bibfnamefont {A.}~\bibnamefont {Zettl}}, \bibinfo
  {author} {\bibfnamefont {M.~L.}\ \bibnamefont {Cohen}}, \ and\ \bibinfo
  {author} {\bibfnamefont {V.~H.}\ \bibnamefont {Crespi}},\ }\href {\doibase
  10.1103/PhysRevLett.83.404} {\bibfield  {journal} {\bibinfo  {journal} {Phys.
  Rev. Lett.}\ }\textbf {\bibinfo {volume} {83}},\ \bibinfo {pages} {404}
  (\bibinfo {year} {1999})}\BibitemShut {NoStop}%
\bibitem [{\citenamefont {Hinks}\ \emph
  {et~al.}(2001{\natexlab{b}})\citenamefont {Hinks}, \citenamefont {Claus},\
  and\ \citenamefont {Jorgensen}}]{hinks457}%
  \BibitemOpen
  \bibfield  {author} {\bibinfo {author} {\bibfnamefont {D.~G.}\ \bibnamefont
  {Hinks}}, \bibinfo {author} {\bibfnamefont {H.}~\bibnamefont {Claus}}, \ and\
  \bibinfo {author} {\bibfnamefont {J.~D.}\ \bibnamefont {Jorgensen}},\ }\href
  {\doibase http://dx.doi.org/10.1038/35078037} {\bibfield  {journal} {\bibinfo
   {journal} {Nature}\ }\textbf {\bibinfo {volume} {457}},\ \bibinfo {pages}
  {457} (\bibinfo {year} {2001}{\natexlab{b}})}\BibitemShut {NoStop}%
\bibitem [{\citenamefont {Batlogg}\ \emph {et~al.}(1987)\citenamefont
  {Batlogg}, \citenamefont {Cava}, \citenamefont {Jayaraman}, \citenamefont
  {van Dover}, \citenamefont {Kourouklis}, \citenamefont {Sunshine},
  \citenamefont {Murphy}, \citenamefont {Rupp}, \citenamefont {Chen},
  \citenamefont {White}, \citenamefont {Short}, \citenamefont {Mujsce},\ and\
  \citenamefont {Rietman}}]{PhysRevLett.58.2333}%
  \BibitemOpen
  \bibfield  {author} {\bibinfo {author} {\bibfnamefont {B.}~\bibnamefont
  {Batlogg}}, \bibinfo {author} {\bibfnamefont {R.~J.}\ \bibnamefont {Cava}},
  \bibinfo {author} {\bibfnamefont {A.}~\bibnamefont {Jayaraman}}, \bibinfo
  {author} {\bibfnamefont {R.~B.}\ \bibnamefont {van Dover}}, \bibinfo {author}
  {\bibfnamefont {G.~A.}\ \bibnamefont {Kourouklis}}, \bibinfo {author}
  {\bibfnamefont {S.}~\bibnamefont {Sunshine}}, \bibinfo {author}
  {\bibfnamefont {D.~W.}\ \bibnamefont {Murphy}}, \bibinfo {author}
  {\bibfnamefont {L.~W.}\ \bibnamefont {Rupp}}, \bibinfo {author}
  {\bibfnamefont {H.~S.}\ \bibnamefont {Chen}}, \bibinfo {author}
  {\bibfnamefont {A.}~\bibnamefont {White}}, \bibinfo {author} {\bibfnamefont
  {K.~T.}\ \bibnamefont {Short}}, \bibinfo {author} {\bibfnamefont {A.~M.}\
  \bibnamefont {Mujsce}}, \ and\ \bibinfo {author} {\bibfnamefont {E.~A.}\
  \bibnamefont {Rietman}},\ }\href {\doibase 10.1103/PhysRevLett.58.2333}
  {\bibfield  {journal} {\bibinfo  {journal} {Phys. Rev. Lett.}\ }\textbf
  {\bibinfo {volume} {58}},\ \bibinfo {pages} {2333} (\bibinfo {year}
  {1987})}\BibitemShut {NoStop}%
\bibitem [{\citenamefont {Liu}\ \emph {et~al.}(2009)\citenamefont {Liu},
  \citenamefont {Wu}, \citenamefont {Wu}, \citenamefont {Chen}, \citenamefont
  {Wang}, \citenamefont {Xie}, \citenamefont {Ying}, \citenamefont {Yan},
  \citenamefont {Li}, \citenamefont {Shi}, \citenamefont {Chu}, \citenamefont
  {Wu},\ and\ \citenamefont {Chen}}]{liu64}%
  \BibitemOpen
  \bibfield  {author} {\bibinfo {author} {\bibfnamefont {R.~H.}\ \bibnamefont
  {Liu}}, \bibinfo {author} {\bibfnamefont {T.}~\bibnamefont {Wu}}, \bibinfo
  {author} {\bibfnamefont {G.}~\bibnamefont {Wu}}, \bibinfo {author}
  {\bibfnamefont {H.}~\bibnamefont {Chen}}, \bibinfo {author} {\bibfnamefont
  {X.~F.}\ \bibnamefont {Wang}}, \bibinfo {author} {\bibfnamefont {Y.~L.}\
  \bibnamefont {Xie}}, \bibinfo {author} {\bibfnamefont {J.~J.}\ \bibnamefont
  {Ying}}, \bibinfo {author} {\bibfnamefont {Y.~J.}\ \bibnamefont {Yan}},
  \bibinfo {author} {\bibfnamefont {Q.~J.}\ \bibnamefont {Li}}, \bibinfo
  {author} {\bibfnamefont {B.~C.}\ \bibnamefont {Shi}}, \bibinfo {author}
  {\bibfnamefont {W.~S.}\ \bibnamefont {Chu}}, \bibinfo {author} {\bibfnamefont
  {Z.~Y.}\ \bibnamefont {Wu}}, \ and\ \bibinfo {author} {\bibfnamefont {X.~H.}\
  \bibnamefont {Chen}},\ }\href {\doibase
  http://dx.doi.org/10.1038/nature07981} {\bibfield  {journal} {\bibinfo
  {journal} {Nature}\ }\textbf {\bibinfo {volume} {459}},\ \bibinfo {pages}
  {64} (\bibinfo {year} {2009})}\BibitemShut {NoStop}%
\bibitem [{\citenamefont {Stritzker}\ and\ \citenamefont
  {Buckel}(1972)}]{stritzker257}%
  \BibitemOpen
  \bibfield  {author} {\bibinfo {author} {\bibfnamefont {B.}~\bibnamefont
  {Stritzker}}\ and\ \bibinfo {author} {\bibfnamefont {W.}~\bibnamefont
  {Buckel}},\ }\href {\doibase 10.1007/BF01398191} {\bibfield  {journal}
  {\bibinfo  {journal} {Zeitschrift für Physik}\ }\textbf {\bibinfo {volume}
  {257}},\ \bibinfo {pages} {1} (\bibinfo {year} {1972})}\BibitemShut {NoStop}%
\bibitem [{\citenamefont {Schirber}\ and\ \citenamefont
  {Northrup}(1974)}]{PhysRevB.10.3818}%
  \BibitemOpen
  \bibfield  {author} {\bibinfo {author} {\bibfnamefont {J.~E.}\ \bibnamefont
  {Schirber}}\ and\ \bibinfo {author} {\bibfnamefont {C.~J.~M.}\ \bibnamefont
  {Northrup}},\ }\href {\doibase 10.1103/PhysRevB.10.3818} {\bibfield
  {journal} {\bibinfo  {journal} {Phys. Rev. B}\ }\textbf {\bibinfo {volume}
  {10}},\ \bibinfo {pages} {3818} (\bibinfo {year} {1974})}\BibitemShut
  {NoStop}%
\bibitem [{\citenamefont {Schirber}\ \emph {et~al.}(1984)\citenamefont
  {Schirber}, \citenamefont {Mintz},\ and\ \citenamefont
  {Wall}}]{Schirber1984837}%
  \BibitemOpen
  \bibfield  {author} {\bibinfo {author} {\bibfnamefont {J.~E.}\ \bibnamefont
  {Schirber}}, \bibinfo {author} {\bibfnamefont {J.~M.}\ \bibnamefont {Mintz}},
  \ and\ \bibinfo {author} {\bibfnamefont {W.}~\bibnamefont {Wall}},\ }\href
  {\doibase 10.1016/0038-1098(84)90251-5} {\bibfield  {journal} {\bibinfo
  {journal} {Solid State Communications}\ }\textbf {\bibinfo {volume} {52}},\
  \bibinfo {pages} {837 } (\bibinfo {year} {1984})}\BibitemShut {NoStop}%
\bibitem [{\citenamefont {Karakozov}\ and\ \citenamefont
  {Maksimov}(1978)}]{karakozov681}%
  \BibitemOpen
  \bibfield  {author} {\bibinfo {author} {\bibfnamefont {A.~E.}\ \bibnamefont
  {Karakozov}}\ and\ \bibinfo {author} {\bibfnamefont {E.~G.}\ \bibnamefont
  {Maksimov}},\ }\href
  {http://www.jetp.ac.ru/cgi-bin/e/index/e/47/2/p358?a=list} {\bibfield
  {journal} {\bibinfo  {journal} {Zh. Eksp. Teor. Fiz.}\ }\textbf {\bibinfo
  {volume} {74}},\ \bibinfo {pages} {681} (\bibinfo {year} {1978})}\BibitemShut
  {NoStop}%
\bibitem [{\citenamefont {Kolesnikov}\ \emph {et~al.}(1991)\citenamefont
  {Kolesnikov}, \citenamefont {Natkaniec}, \citenamefont {Antonov},
  \citenamefont {Belash}, \citenamefont {Fedotov}, \citenamefont {Krawczyk},
  \citenamefont {Mayer},\ and\ \citenamefont
  {Ponyatovsky}}]{Kolesnikov1991257}%
  \BibitemOpen
  \bibfield  {author} {\bibinfo {author} {\bibfnamefont {A.}~\bibnamefont
  {Kolesnikov}}, \bibinfo {author} {\bibfnamefont {I.}~\bibnamefont
  {Natkaniec}}, \bibinfo {author} {\bibfnamefont {V.}~\bibnamefont {Antonov}},
  \bibinfo {author} {\bibfnamefont {I.}~\bibnamefont {Belash}}, \bibinfo
  {author} {\bibfnamefont {V.}~\bibnamefont {Fedotov}}, \bibinfo {author}
  {\bibfnamefont {J.}~\bibnamefont {Krawczyk}}, \bibinfo {author}
  {\bibfnamefont {J.}~\bibnamefont {Mayer}}, \ and\ \bibinfo {author}
  {\bibfnamefont {E.}~\bibnamefont {Ponyatovsky}},\ }\href {\doibase
  10.1016/0921-4526(91)90616-M} {\bibfield  {journal} {\bibinfo  {journal}
  {Physica B: Condensed Matter}\ }\textbf {\bibinfo {volume} {174}},\ \bibinfo
  {pages} {257 } (\bibinfo {year} {1991})}\BibitemShut {NoStop}%
\bibitem [{\citenamefont {Sherman}\ \emph {et~al.}(1977)\citenamefont
  {Sherman}, \citenamefont {Birnbaum}, \citenamefont {Holy},\ and\
  \citenamefont {Klein}}]{Sherman1977353}%
  \BibitemOpen
  \bibfield  {author} {\bibinfo {author} {\bibfnamefont {R.}~\bibnamefont
  {Sherman}}, \bibinfo {author} {\bibfnamefont {H.~K.}\ \bibnamefont
  {Birnbaum}}, \bibinfo {author} {\bibfnamefont {J.~A.}\ \bibnamefont {Holy}},
  \ and\ \bibinfo {author} {\bibfnamefont {M.~V.}\ \bibnamefont {Klein}},\
  }\href {\doibase 10.1016/0375-9601(77)90439-X} {\bibfield  {journal}
  {\bibinfo  {journal} {Physics Letters A}\ }\textbf {\bibinfo {volume} {62}},\
  \bibinfo {pages} {353 } (\bibinfo {year} {1977})}\BibitemShut {NoStop}%
\bibitem [{\citenamefont {Ross}\ \emph {et~al.}(1998)\citenamefont {Ross},
  \citenamefont {Antonov}, \citenamefont {Bokhenkov}, \citenamefont
  {Kolesnikov}, \citenamefont {Ponyatovsky},\ and\ \citenamefont
  {Tomkinson}}]{PhysRevB.58.2591}%
  \BibitemOpen
  \bibfield  {author} {\bibinfo {author} {\bibfnamefont {D.~K.}\ \bibnamefont
  {Ross}}, \bibinfo {author} {\bibfnamefont {V.~E.}\ \bibnamefont {Antonov}},
  \bibinfo {author} {\bibfnamefont {E.~L.}\ \bibnamefont {Bokhenkov}}, \bibinfo
  {author} {\bibfnamefont {A.~I.}\ \bibnamefont {Kolesnikov}}, \bibinfo
  {author} {\bibfnamefont {E.~G.}\ \bibnamefont {Ponyatovsky}}, \ and\ \bibinfo
  {author} {\bibfnamefont {J.}~\bibnamefont {Tomkinson}},\ }\href {\doibase
  10.1103/PhysRevB.58.2591} {\bibfield  {journal} {\bibinfo  {journal} {Phys.
  Rev. B}\ }\textbf {\bibinfo {volume} {58}},\ \bibinfo {pages} {2591}
  (\bibinfo {year} {1998})}\BibitemShut {NoStop}%
\bibitem [{\citenamefont {Chowdhury}\ and\ \citenamefont
  {Ross}(1973)}]{Chowdhury1973229}%
  \BibitemOpen
  \bibfield  {author} {\bibinfo {author} {\bibfnamefont {M.}~\bibnamefont
  {Chowdhury}}\ and\ \bibinfo {author} {\bibfnamefont {D.}~\bibnamefont
  {Ross}},\ }\href {\doibase 10.1016/0038-1098(73)90231-7} {\bibfield
  {journal} {\bibinfo  {journal} {Solid State Communications}\ }\textbf
  {\bibinfo {volume} {13}},\ \bibinfo {pages} {229 } (\bibinfo {year}
  {1973})}\BibitemShut {NoStop}%
\bibitem [{\citenamefont {Rowe}\ \emph {et~al.}(1974)\citenamefont {Rowe},
  \citenamefont {Rush}, \citenamefont {Smith}, \citenamefont {Mostoller},\ and\
  \citenamefont {Flotow}}]{PhysRevLett.33.1297}%
  \BibitemOpen
  \bibfield  {author} {\bibinfo {author} {\bibfnamefont {J.~M.}\ \bibnamefont
  {Rowe}}, \bibinfo {author} {\bibfnamefont {J.~J.}\ \bibnamefont {Rush}},
  \bibinfo {author} {\bibfnamefont {H.~G.}\ \bibnamefont {Smith}}, \bibinfo
  {author} {\bibfnamefont {M.}~\bibnamefont {Mostoller}}, \ and\ \bibinfo
  {author} {\bibfnamefont {H.~E.}\ \bibnamefont {Flotow}},\ }\href {\doibase
  10.1103/PhysRevLett.33.1297} {\bibfield  {journal} {\bibinfo  {journal}
  {Phys. Rev. Lett.}\ }\textbf {\bibinfo {volume} {33}},\ \bibinfo {pages}
  {1297} (\bibinfo {year} {1974})}\BibitemShut {NoStop}%
\bibitem [{\citenamefont {Rowe}\ \emph {et~al.}(1986)\citenamefont {Rowe},
  \citenamefont {Rush}, \citenamefont {Schirber},\ and\ \citenamefont
  {Mintz}}]{PhysRevLett.57.2955}%
  \BibitemOpen
  \bibfield  {author} {\bibinfo {author} {\bibfnamefont {J.~M.}\ \bibnamefont
  {Rowe}}, \bibinfo {author} {\bibfnamefont {J.~J.}\ \bibnamefont {Rush}},
  \bibinfo {author} {\bibfnamefont {J.~E.}\ \bibnamefont {Schirber}}, \ and\
  \bibinfo {author} {\bibfnamefont {J.~M.}\ \bibnamefont {Mintz}},\ }\href
  {\doibase 10.1103/PhysRevLett.57.2955} {\bibfield  {journal} {\bibinfo
  {journal} {Phys. Rev. Lett.}\ }\textbf {\bibinfo {volume} {57}},\ \bibinfo
  {pages} {2955} (\bibinfo {year} {1986})}\BibitemShut {NoStop}%
\bibitem [{\citenamefont {Miller}\ and\ \citenamefont
  {Satterthwaite}(1975)}]{PhysRevLett.34.144}%
  \BibitemOpen
  \bibfield  {author} {\bibinfo {author} {\bibfnamefont {R.~J.}\ \bibnamefont
  {Miller}}\ and\ \bibinfo {author} {\bibfnamefont {C.~B.}\ \bibnamefont
  {Satterthwaite}},\ }\href {\doibase 10.1103/PhysRevLett.34.144} {\bibfield
  {journal} {\bibinfo  {journal} {Phys. Rev. Lett.}\ }\textbf {\bibinfo
  {volume} {34}},\ \bibinfo {pages} {144} (\bibinfo {year} {1975})}\BibitemShut
  {NoStop}%
\bibitem [{\citenamefont {Klein}\ \emph {et~al.}(1977)\citenamefont {Klein},
  \citenamefont {Economou},\ and\ \citenamefont
  {Papaconstantopoulos}}]{PhysRevLett.39.574}%
  \BibitemOpen
  \bibfield  {author} {\bibinfo {author} {\bibfnamefont {B.~M.}\ \bibnamefont
  {Klein}}, \bibinfo {author} {\bibfnamefont {E.~N.}\ \bibnamefont {Economou}},
  \ and\ \bibinfo {author} {\bibfnamefont {D.~A.}\ \bibnamefont
  {Papaconstantopoulos}},\ }\href {\doibase 10.1103/PhysRevLett.39.574}
  {\bibfield  {journal} {\bibinfo  {journal} {Phys. Rev. Lett.}\ }\textbf
  {\bibinfo {volume} {39}},\ \bibinfo {pages} {574} (\bibinfo {year}
  {1977})}\BibitemShut {NoStop}%
\bibitem [{\citenamefont {Els\"asser}\ \emph {et~al.}(1991)\citenamefont
  {Els\"asser}, \citenamefont {Ho}, \citenamefont {Chan},\ and\ \citenamefont
  {F\"ahnle}}]{PhysRevB.44.10377}%
  \BibitemOpen
  \bibfield  {author} {\bibinfo {author} {\bibfnamefont {C.}~\bibnamefont
  {Els\"asser}}, \bibinfo {author} {\bibfnamefont {K.~M.}\ \bibnamefont {Ho}},
  \bibinfo {author} {\bibfnamefont {C.~T.}\ \bibnamefont {Chan}}, \ and\
  \bibinfo {author} {\bibfnamefont {M.}~\bibnamefont {F\"ahnle}},\ }\href
  {\doibase 10.1103/PhysRevB.44.10377} {\bibfield  {journal} {\bibinfo
  {journal} {Phys. Rev. B}\ }\textbf {\bibinfo {volume} {44}},\ \bibinfo
  {pages} {10377} (\bibinfo {year} {1991})}\BibitemShut {NoStop}%
\bibitem [{\citenamefont {Klein}\ and\ \citenamefont
  {Cohen}(1992)}]{PhysRevB.45.12405}%
  \BibitemOpen
  \bibfield  {author} {\bibinfo {author} {\bibfnamefont {B.~M.}\ \bibnamefont
  {Klein}}\ and\ \bibinfo {author} {\bibfnamefont {R.~E.}\ \bibnamefont
  {Cohen}},\ }\href {\doibase 10.1103/PhysRevB.45.12405} {\bibfield  {journal}
  {\bibinfo  {journal} {Phys. Rev. B}\ }\textbf {\bibinfo {volume} {45}},\
  \bibinfo {pages} {12405} (\bibinfo {year} {1992})}\BibitemShut {NoStop}%
\bibitem [{\citenamefont {Schirber}\ and\ \citenamefont
  {Morosin}(1975)}]{PhysRevB.12.117}%
  \BibitemOpen
  \bibfield  {author} {\bibinfo {author} {\bibfnamefont {J.~E.}\ \bibnamefont
  {Schirber}}\ and\ \bibinfo {author} {\bibfnamefont {B.}~\bibnamefont
  {Morosin}},\ }\href {\doibase 10.1103/PhysRevB.12.117} {\bibfield  {journal}
  {\bibinfo  {journal} {Phys. Rev. B}\ }\textbf {\bibinfo {volume} {12}},\
  \bibinfo {pages} {117} (\bibinfo {year} {1975})}\BibitemShut {NoStop}%
\bibitem [{\citenamefont {Jena}\ \emph {et~al.}(1984)\citenamefont {Jena},
  \citenamefont {Jones},\ and\ \citenamefont {Nieminen}}]{PhysRevB.29.4140}%
  \BibitemOpen
  \bibfield  {author} {\bibinfo {author} {\bibfnamefont {P.}~\bibnamefont
  {Jena}}, \bibinfo {author} {\bibfnamefont {J.}~\bibnamefont {Jones}}, \ and\
  \bibinfo {author} {\bibfnamefont {R.~M.}\ \bibnamefont {Nieminen}},\ }\href
  {\doibase 10.1103/PhysRevB.29.4140} {\bibfield  {journal} {\bibinfo
  {journal} {Phys. Rev. B}\ }\textbf {\bibinfo {volume} {29}},\ \bibinfo
  {pages} {4140} (\bibinfo {year} {1984})}\BibitemShut {NoStop}%
\bibitem [{\citenamefont {Rousseau}\ and\ \citenamefont
  {Bergara}(2010)}]{PhysRevB.82.104504}%
  \BibitemOpen
  \bibfield  {author} {\bibinfo {author} {\bibfnamefont {B.}~\bibnamefont
  {Rousseau}}\ and\ \bibinfo {author} {\bibfnamefont {A.}~\bibnamefont
  {Bergara}},\ }\href {\doibase 10.1103/PhysRevB.82.104504} {\bibfield
  {journal} {\bibinfo  {journal} {Phys. Rev. B}\ }\textbf {\bibinfo {volume}
  {82}},\ \bibinfo {pages} {104504} (\bibinfo {year} {2010})}\BibitemShut
  {NoStop}%
\bibitem [{\citenamefont {Paulatto}\ \emph {et~al.}(2013)\citenamefont
  {Paulatto}, \citenamefont {Mauri},\ and\ \citenamefont
  {Lazzeri}}]{PhysRevB.87.214303}%
  \BibitemOpen
  \bibfield  {author} {\bibinfo {author} {\bibfnamefont {L.}~\bibnamefont
  {Paulatto}}, \bibinfo {author} {\bibfnamefont {F.}~\bibnamefont {Mauri}}, \
  and\ \bibinfo {author} {\bibfnamefont {M.}~\bibnamefont {Lazzeri}},\ }\href
  {\doibase 10.1103/PhysRevB.87.214303} {\bibfield  {journal} {\bibinfo
  {journal} {Phys. Rev. B}\ }\textbf {\bibinfo {volume} {87}},\ \bibinfo
  {pages} {214303} (\bibinfo {year} {2013})}\BibitemShut {NoStop}%
\bibitem [{\citenamefont {Lazzeri}\ and\ \citenamefont
  {de~Gironcoli}(2002)}]{Lazzeri}%
  \BibitemOpen
  \bibfield  {author} {\bibinfo {author} {\bibfnamefont {M.}~\bibnamefont
  {Lazzeri}}\ and\ \bibinfo {author} {\bibfnamefont {S.}~\bibnamefont
  {de~Gironcoli}},\ }\href {\doibase 10.1103/PhysRevB.65.245402} {\bibfield
  {journal} {\bibinfo  {journal} {Phys. Rev. B}\ }\textbf {\bibinfo {volume}
  {65}},\ \bibinfo {pages} {245402} (\bibinfo {year} {2002})}\BibitemShut
  {NoStop}%
\bibitem [{\citenamefont {Hooton}(1955)}]{hooton422}%
  \BibitemOpen
  \bibfield  {author} {\bibinfo {author} {\bibfnamefont {D.~J.}\ \bibnamefont
  {Hooton}},\ }\href@noop {} {\bibfield  {journal} {\bibinfo  {journal}
  {Philosophical Magazine Series 7}\ }\textbf {\bibinfo {volume} {46}},\
  \bibinfo {pages} {422} (\bibinfo {year} {1955})}\BibitemShut {NoStop}%
\bibitem [{\citenamefont {Souvatzis}\ \emph {et~al.}(2008)\citenamefont
  {Souvatzis}, \citenamefont {Eriksson}, \citenamefont {Katsnelson},\ and\
  \citenamefont {Rudin}}]{PhysRevLett.100.095901}%
  \BibitemOpen
  \bibfield  {author} {\bibinfo {author} {\bibfnamefont {P.}~\bibnamefont
  {Souvatzis}}, \bibinfo {author} {\bibfnamefont {O.}~\bibnamefont {Eriksson}},
  \bibinfo {author} {\bibfnamefont {M.~I.}\ \bibnamefont {Katsnelson}}, \ and\
  \bibinfo {author} {\bibfnamefont {S.~P.}\ \bibnamefont {Rudin}},\ }\href
  {\doibase 10.1103/PhysRevLett.100.095901} {\bibfield  {journal} {\bibinfo
  {journal} {Phys. Rev. Lett.}\ }\textbf {\bibinfo {volume} {100}},\ \bibinfo
  {pages} {095901} (\bibinfo {year} {2008})}\BibitemShut {NoStop}%
\bibitem [{\citenamefont {Hellman}\ \emph {et~al.}(2011)\citenamefont
  {Hellman}, \citenamefont {Abrikosov},\ and\ \citenamefont
  {Simak}}]{PhysRevB.84.180301}%
  \BibitemOpen
  \bibfield  {author} {\bibinfo {author} {\bibfnamefont {O.}~\bibnamefont
  {Hellman}}, \bibinfo {author} {\bibfnamefont {I.~A.}\ \bibnamefont
  {Abrikosov}}, \ and\ \bibinfo {author} {\bibfnamefont {S.~I.}\ \bibnamefont
  {Simak}},\ }\href {\doibase 10.1103/PhysRevB.84.180301} {\bibfield  {journal}
  {\bibinfo  {journal} {Phys. Rev. B}\ }\textbf {\bibinfo {volume} {84}},\
  \bibinfo {pages} {180301} (\bibinfo {year} {2011})}\BibitemShut {NoStop}%
\bibitem [{\citenamefont {Monserrat}\ \emph {et~al.}(2013)\citenamefont
  {Monserrat}, \citenamefont {Drummond},\ and\ \citenamefont {Needs}}]{Needs}%
  \BibitemOpen
  \bibfield  {author} {\bibinfo {author} {\bibfnamefont {B.}~\bibnamefont
  {Monserrat}}, \bibinfo {author} {\bibfnamefont {N.~D.}\ \bibnamefont
  {Drummond}}, \ and\ \bibinfo {author} {\bibfnamefont {R.~J.}\ \bibnamefont
  {Needs}},\ }\href {\doibase 10.1103/PhysRevB.87.144302} {\bibfield  {journal}
  {\bibinfo  {journal} {Phys. Rev. B}\ }\textbf {\bibinfo {volume} {87}},\
  \bibinfo {pages} {144302} (\bibinfo {year} {2013})}\BibitemShut {NoStop}%
\bibitem [{\citenamefont {Errea}\ \emph {et~al.}(2011)\citenamefont {Errea},
  \citenamefont {Rousseau},\ and\ \citenamefont
  {Bergara}}]{PhysRevLett.106.165501}%
  \BibitemOpen
  \bibfield  {author} {\bibinfo {author} {\bibfnamefont {I.}~\bibnamefont
  {Errea}}, \bibinfo {author} {\bibfnamefont {B.}~\bibnamefont {Rousseau}}, \
  and\ \bibinfo {author} {\bibfnamefont {A.}~\bibnamefont {Bergara}},\ }\href
  {\doibase 10.1103/PhysRevLett.106.165501} {\bibfield  {journal} {\bibinfo
  {journal} {Phys. Rev. Lett.}\ }\textbf {\bibinfo {volume} {106}},\ \bibinfo
  {pages} {165501} (\bibinfo {year} {2011})}\BibitemShut {NoStop}%
\bibitem [{\citenamefont {Errea}\ \emph {et~al.}(2012)\citenamefont {Errea},
  \citenamefont {Rousseau},\ and\ \citenamefont {Bergara}}]{errea:112604}%
  \BibitemOpen
  \bibfield  {author} {\bibinfo {author} {\bibfnamefont {I.}~\bibnamefont
  {Errea}}, \bibinfo {author} {\bibfnamefont {B.}~\bibnamefont {Rousseau}}, \
  and\ \bibinfo {author} {\bibfnamefont {A.}~\bibnamefont {Bergara}},\ }\href
  {\doibase 10.1063/1.4726161} {\bibfield  {journal} {\bibinfo  {journal}
  {Journal of Applied Physics}\ }\textbf {\bibinfo {volume} {111}},\ \bibinfo
  {eid} {112604} (\bibinfo {year} {2012})}\BibitemShut {NoStop}%
\bibitem [{\citenamefont {Rossano}\ \emph {et~al.}(2005)\citenamefont
  {Rossano}, \citenamefont {Mauri}, \citenamefont {Pickard},\ and\
  \citenamefont {Farnan}}]{doi:10.1021/jp044251w}%
  \BibitemOpen
  \bibfield  {author} {\bibinfo {author} {\bibfnamefont {S.}~\bibnamefont
  {Rossano}}, \bibinfo {author} {\bibfnamefont {F.}~\bibnamefont {Mauri}},
  \bibinfo {author} {\bibfnamefont {C.~J.}\ \bibnamefont {Pickard}}, \ and\
  \bibinfo {author} {\bibfnamefont {I.}~\bibnamefont {Farnan}},\ }\href
  {\doibase 10.1021/jp044251w} {\bibfield  {journal} {\bibinfo  {journal} {The
  Journal of Physical Chemistry B}\ }\textbf {\bibinfo {volume} {109}},\
  \bibinfo {pages} {7245} (\bibinfo {year} {2005})}\BibitemShut {NoStop}%
\bibitem [{\citenamefont {Abbenseth}\ and\ \citenamefont
  {Wipf}(1980)}]{0305-4608-10-3-006}%
  \BibitemOpen
  \bibfield  {author} {\bibinfo {author} {\bibfnamefont {R.}~\bibnamefont
  {Abbenseth}}\ and\ \bibinfo {author} {\bibfnamefont {H.}~\bibnamefont
  {Wipf}},\ }\href {http://stacks.iop.org/0305-4608/10/i=3/a=006} {\bibfield
  {journal} {\bibinfo  {journal} {Journal of Physics F: Metal Physics}\
  }\textbf {\bibinfo {volume} {10}},\ \bibinfo {pages} {353} (\bibinfo {year}
  {1980})}\BibitemShut {NoStop}%
\bibitem [{\citenamefont {Giannozzi~\emph{et
  al.}}(2009)}]{0953-8984-21-39-395502}%
  \BibitemOpen
  \bibfield  {author} {\bibinfo {author} {\bibfnamefont {P.}~\bibnamefont
  {Giannozzi~\emph{et al.}}},\ }\href@noop {} {\bibfield  {journal} {\bibinfo
  {journal} {J. Phys. Condens. Matter}\ }\textbf {\bibinfo {volume} {21}},\
  \bibinfo {pages} {395502} (\bibinfo {year} {2009})}\BibitemShut {NoStop}%
\bibitem [{\citenamefont {Baroni}\ \emph {et~al.}(2001)\citenamefont {Baroni},
  \citenamefont {de~Gironcoli}, \citenamefont {Dal~Corso},\ and\ \citenamefont
  {Giannozzi}}]{RevModPhys.73.515}%
  \BibitemOpen
  \bibfield  {author} {\bibinfo {author} {\bibfnamefont {S.}~\bibnamefont
  {Baroni}}, \bibinfo {author} {\bibfnamefont {S.}~\bibnamefont
  {de~Gironcoli}}, \bibinfo {author} {\bibfnamefont {A.}~\bibnamefont
  {Dal~Corso}}, \ and\ \bibinfo {author} {\bibfnamefont {P.}~\bibnamefont
  {Giannozzi}},\ }\href {\doibase 10.1103/RevModPhys.73.515} {\bibfield
  {journal} {\bibinfo  {journal} {Rev. Mod. Phys.}\ }\textbf {\bibinfo {volume}
  {73}},\ \bibinfo {pages} {515} (\bibinfo {year} {2001})}\BibitemShut
  {NoStop}%
\bibitem [{\citenamefont {Perdew}\ and\ \citenamefont
  {Zunger}(1981)}]{PhysRevB.23.5048}%
  \BibitemOpen
  \bibfield  {author} {\bibinfo {author} {\bibfnamefont {J.~P.}\ \bibnamefont
  {Perdew}}\ and\ \bibinfo {author} {\bibfnamefont {A.}~\bibnamefont
  {Zunger}},\ }\href {\doibase 10.1103/PhysRevB.23.5048} {\bibfield  {journal}
  {\bibinfo  {journal} {Phys. Rev. B}\ }\textbf {\bibinfo {volume} {23}},\
  \bibinfo {pages} {5048} (\bibinfo {year} {1981})}\BibitemShut {NoStop}%
\bibitem [{Not()}]{Note1}%
  \BibitemOpen
  \href@noop {} {}\bibinfo {note} {Calculations are performed applying the {\sc
  Quantum-ESPRESSO}~\cite {0953-8984-21-39-395502} code. A 50 Ry cutoff is used
  for the plane-wave basis and a 24 $\times $ 24 $\times $ 24 mesh for the BZ
  integrations in the unit cell. The sum over $\protect \mathbf {k}$ in Eq.
  \eqref{eliashberg} required a 72 $\times $ 72 $\times $ 72 grid.}\BibitemShut
  {Stop}%
\bibitem [{foo()}]{footnote}%
  \BibitemOpen
  \href@noop {} {}\bibinfo {note} {Total energies and forces needed in Eqs.
  \eqref{potentialintegral} and \eqref{gradient} are computed in a 2 $\times$ 2
  $\times$ 2 supercell containing 16 atoms. The difference between the SSCHA
  force constant matrix and the harmonic force constant matrix in the 2
  $\times$ 2 $\times$ 2 supercell is interpolated to a 4 $\times$ 4 $\times$ 4
  supercell. The harmonic 4 $\times$ 4 $\times$ 4 force constant matrix is
  added to the result. We verified that the anharmonic phonon dispersion
  obtained for the model potential in this way coincides with that obtained
  directly in a 4 $\times$ 4 $\times$ 4 supercell. We use $N_c = 20000$ and
  $N_c = 300$ for the calculations with the model and \abinitio{} potential,
  respectively. For the \abinitio{} calculation a single set
  $\{\mathbf{R}_I\}_{I=1,\dots,N_c}$ is sufficient if the result of the model
  potential is used as ${\mathcal{H}}_0$.}\BibitemShut {Stop}%
\bibitem [{\citenamefont {Zhong}\ \emph {et~al.}(1995)\citenamefont {Zhong},
  \citenamefont {Vanderbilt},\ and\ \citenamefont {Rabe}}]{PhysRevB.52.6301}%
  \BibitemOpen
  \bibfield  {author} {\bibinfo {author} {\bibfnamefont {W.}~\bibnamefont
  {Zhong}}, \bibinfo {author} {\bibfnamefont {D.}~\bibnamefont {Vanderbilt}}, \
  and\ \bibinfo {author} {\bibfnamefont {K.~M.}\ \bibnamefont {Rabe}},\ }\href
  {\doibase 10.1103/PhysRevB.52.6301} {\bibfield  {journal} {\bibinfo
  {journal} {Phys. Rev. B}\ }\textbf {\bibinfo {volume} {52}},\ \bibinfo
  {pages} {6301} (\bibinfo {year} {1995})}\BibitemShut {NoStop}%
\bibitem [{\citenamefont {Allen}\ and\ \citenamefont
  {Dynes}(1975)}]{PhysRevB.12.905}%
  \BibitemOpen
  \bibfield  {author} {\bibinfo {author} {\bibfnamefont {P.~B.}\ \bibnamefont
  {Allen}}\ and\ \bibinfo {author} {\bibfnamefont {R.~C.}\ \bibnamefont
  {Dynes}},\ }\href {\doibase 10.1103/PhysRevB.12.905} {\bibfield  {journal}
  {\bibinfo  {journal} {Phys. Rev. B}\ }\textbf {\bibinfo {volume} {12}},\
  \bibinfo {pages} {905} (\bibinfo {year} {1975})}\BibitemShut {NoStop}%
\bibitem [{\citenamefont {Kim}\ \emph {et~al.}(2010)\citenamefont {Kim},
  \citenamefont {Scheicher}, \citenamefont {Mao}, \citenamefont {Kang},\ and\
  \citenamefont {Ahuja}}]{Kim16022010}%
  \BibitemOpen
  \bibfield  {author} {\bibinfo {author} {\bibfnamefont {D.~Y.}\ \bibnamefont
  {Kim}}, \bibinfo {author} {\bibfnamefont {R.~H.}\ \bibnamefont {Scheicher}},
  \bibinfo {author} {\bibfnamefont {H.-k.}\ \bibnamefont {Mao}}, \bibinfo
  {author} {\bibfnamefont {T.~W.}\ \bibnamefont {Kang}}, \ and\ \bibinfo
  {author} {\bibfnamefont {R.}~\bibnamefont {Ahuja}},\ }\href {\doibase
  10.1073/pnas.0914462107} {\bibfield  {journal} {\bibinfo  {journal} {Proc.
  Natl. Acad. Sci. USA}\ }\textbf {\bibinfo {volume} {107}},\ \bibinfo {pages}
  {2793} (\bibinfo {year} {2010})}\BibitemShut {NoStop}%
\bibitem [{\citenamefont {Kim}\ \emph {et~al.}(2011)\citenamefont {Kim},
  \citenamefont {Scheicher}, \citenamefont {Pickard}, \citenamefont {Needs},\
  and\ \citenamefont {Ahuja}}]{PhysRevLett.107.117002}%
  \BibitemOpen
  \bibfield  {author} {\bibinfo {author} {\bibfnamefont {D.~Y.}\ \bibnamefont
  {Kim}}, \bibinfo {author} {\bibfnamefont {R.~H.}\ \bibnamefont {Scheicher}},
  \bibinfo {author} {\bibfnamefont {C.~J.}\ \bibnamefont {Pickard}}, \bibinfo
  {author} {\bibfnamefont {R.~J.}\ \bibnamefont {Needs}}, \ and\ \bibinfo
  {author} {\bibfnamefont {R.}~\bibnamefont {Ahuja}},\ }\href {\doibase
  10.1103/PhysRevLett.107.117002} {\bibfield  {journal} {\bibinfo  {journal}
  {Phys. Rev. Lett.}\ }\textbf {\bibinfo {volume} {107}},\ \bibinfo {pages}
  {117002} (\bibinfo {year} {2011})}\BibitemShut {NoStop}%
\bibitem [{\citenamefont {Gr\"uner}(1994)}]{gruner}%
  \BibitemOpen
  \bibfield  {author} {\bibinfo {author} {\bibfnamefont {G.}~\bibnamefont
  {Gr\"uner}},\ }\href@noop {} {\emph {\bibinfo {title} {Density Waves in
  Solids}}}\ (\bibinfo  {publisher} {Addison-Wesley},\ \bibinfo {address}
  {Reading, MA},\ \bibinfo {year} {1994})\BibitemShut {NoStop}%
\end{thebibliography}
%

\end{document}